\documentclass[sigconf,dvipsnames]{acmart}

\usepackage{pgfgantt}
\usepackage{lipsum}
\usepackage{seqsplit}

\AtBeginDocument{%
  }

\setcopyright{acmlicensed}
\copyrightyear{2018}
\acmYear{2018}
\acmDOI{XXXXXXX.XXXXXXX}
\acmConference[Conference acronym 'XX]{Make sure to enter the correct
  conference title from your rights confirmation email}{June 03--05,
  2018}{Woodstock, NY}
\acmISBN{978-1-4503-XXXX-X/2018/06}

\begin{document}

\title{Resilient Multi-Agent Negotiation for Medical Supply Chains: Integrating LLMs and Blockchain for Transparent Coordination}

\author{Mariam ALMutairi}
\email{malmutairi@vt.edu}
\affiliation{%
  \institution{Virginia Tech}
  \city{Alexandria}
  \state{Virgnia}
  \country{USA}
}

\author{Hyungmin Kim}
\email{henrykim9319@vt.edu}
\affiliation{%
  \institution{Virginia Tech}
  \city{Alexandria}
  \state{Virgnia}
  \country{USA}
}

\renewcommand{\shortauthors}{Trovato et al.}

\begin{abstract}
Global health emergencies, such as the COVID-19 pandemic, have exposed critical weaknesses in traditional medical supply chains, including inefficiencies in resource allocation, lack of transparency, and poor adaptability to dynamic disruptions. This paper presents a novel hybrid framework that integrates blockchain technology with a decentralized, large language model (LLM) powered multi-agent negotiation system to enhance the resilience and accountability of medical supply chains during crises. In this system, autonomous agents—representing manufacturers, distributors, and healthcare institutions—engage in structured, context-aware negotiation and decision-making processes facilitated by LLMs, enabling rapid and ethical allocation of scarce medical resources. The off-chain agent layer supports adaptive reasoning and local decision-making, while the on-chain blockchain layer ensures immutable, transparent, and auditable enforcement of decisions via smart contracts. The framework also incorporates a formal cross-layer communication protocol to bridge decentralized negotiation with institutional enforcement. A simulation environment emulating pandemic scenarios evaluates the system’s performance, demonstrating improvements in negotiation efficiency, fairness of allocation, supply chain responsiveness, and auditability. This research contributes an innovative approach that synergizes blockchain trust guarantees with the adaptive intelligence of LLM-driven agents, providing a robust and scalable solution for critical supply chain coordination under uncertainty.
\end{abstract}



\keywords{Do, Not, Us, This, Code, Put, the, Correct, Terms, for,
  Your, Paper}

\received{20 February 2007}
\received[revised]{12 March 2009}
\received[accepted]{5 June 2009}

\maketitle

\section{Introduction}

In recent years, global health crises, such as the COVID-19 pandemic, have revealed significant vulnerabilities in traditional medical supply chains. These supply chains, typically characterized by centralized management, rigid protocols, and fragmented communication channels, have struggled to respond effectively to dynamic and unpredictable demands that arise during public health emergencies. Delays in the allocation of critical medical supplies—such as personal protective equipment, vaccines, and essential medications—have led to avoidable shortages, inequitable distribution, and preventable health consequences. Moreover, the lack of transparent coordination and trust among key stakeholders, including manufacturers, distributors, healthcare providers, and regulatory authorities, has exacerbated inefficiencies, increased operational risks, and undermined public trust in healthcare delivery systems.

The \textbf{centralized and siloed nature of existing supply chains} makes them particularly prone to systemic failures during disruptions. These systems often lack the \textbf{agility to adjust resource allocation dynamically} in response to evolving crisis scenarios. Additionally, traditional decision-making approaches rely heavily on \textbf{manual intervention, pre-programmed rules, and static protocols}, which limit responsiveness and the ability to address complex ethical considerations such as fairness and prioritization under scarcity. Furthermore, \textbf{data integrity, visibility, and accountability} remain persistent challenges, leaving room for \textbf{misallocation, counterfeiting, and loss of critical situational awareness}.

Emerging technologies offer promising avenues to address these systemic shortcomings. \textbf{Blockchain technology} has been widely explored in supply chains to enhance \textbf{data traceability, security, and transparency}, while \textbf{multi-agent systems} and \textbf{large language models (LLMs)} have shown potential in enabling decentralized, adaptive decision-making and negotiation processes. However, existing approaches typically treat these technologies in isolation, failing to fully exploit their complementary strengths. \textbf{Blockchain alone cannot ensure adaptability in dynamic crises}, while \textbf{multi-agent systems without trust mechanisms risk coordination failures and data disputes}.

To address these gaps, this research proposes an innovative \textbf{hybrid framework} that integrates blockchain technology with an \textbf{LLM-powered multi-agent negotiation system}. This dual-layer architecture leverages the \textbf{autonomy and reasoning capabilities of LLM agents} for dynamic, context-sensitive decision-making, while using \textbf{blockchain-based smart contracts to enforce transparency, auditability, and immutable record-keeping}. By combining these technologies, the proposed system seeks to create a \textbf{decentralized, trustworthy, and responsive coordination mechanism} capable of handling the complexities and uncertainties inherent in medical supply chains during crises.

The main contributions of this research are as follows:
\begin{itemize}
    \item \textbf{Decentralized Multi-Agent Negotiation System}: Development of a multi-agent system where autonomous LLM-powered agents represent key stakeholders and negotiate resource allocation dynamically during crises.
    \item \textbf{Blockchain-Based Enforcement Layer}: Integration of a blockchain execution layer that ensures transparent, tamper-proof recording of decisions and actions, reinforcing trust and accountability.
    \item \textbf{Formal Cross-Layer Communication Protocol}: Introduction of a structured protocol that bridges off-chain decision-making with on-chain enforcement, ensuring coherence, verifiability, and modularity.
    \item \textbf{Simulation and Evaluation}: Implementation of a domain-specific simulation environment to evaluate the framework under pandemic-like conditions, assessing system performance in terms of resilience, fairness, throughput, and auditability.
\end{itemize}

This research contributes to the ongoing discourse on the modernization of healthcare supply chains by presenting a \textbf{novel, decentralized, and intelligent approach} that can improve \textbf{responsiveness, fairness, and trust} in the face of global health emergencies.

\section{Overview}
\begin{figure}[htbp]
  \centering
  \includegraphics[width=\columnwidth]{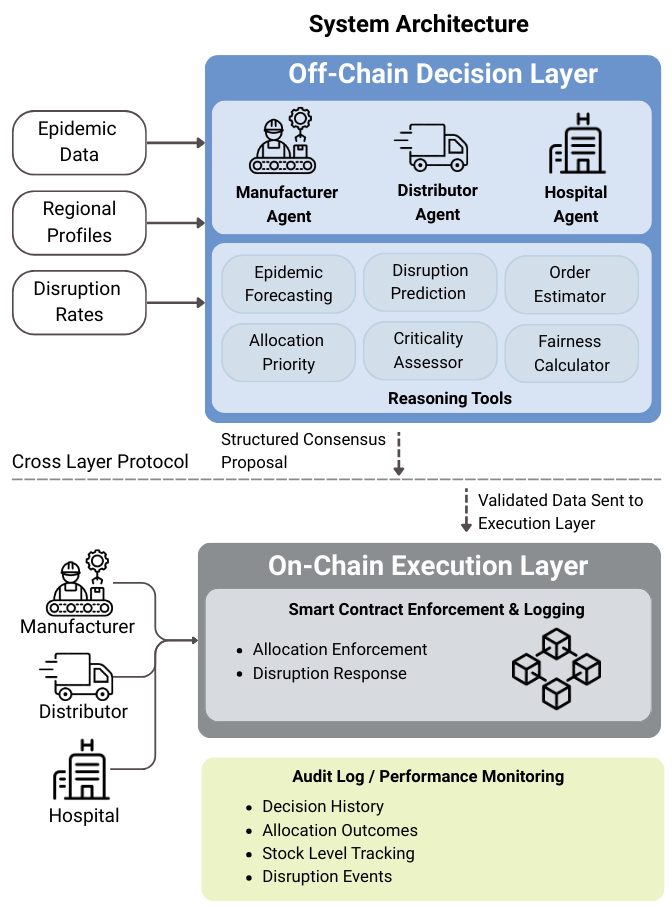} % Update with your actual path
  \caption{Hybrid architecture for decentralized supply chain coordination. The off-chain decision layer uses autonomous agents with embedded reasoning tools, while the on-chain execution layer ensures verifiable enforcement and auditability.}
  \Description{A diagram showing the hybrid architecture for decentralized supply chain coordination, with an off-chain decision layer and an on-chain execution layer.}
  \label{fig:architecture}
\end{figure}

The proposed system implements a dual-layer architecture to facilitate coordinated, high-stakes supply chain decision-making, particularly in the context of public health emergencies such as pandemics (see Figure~\ref{fig:architecture}). This architecture strategically decouples adaptive, context-sensitive decision processes from the execution and record-keeping layer, which emphasizes immutability and verifiability. A well-defined interface enables seamless interaction between the two layers, ensuring coherence and traceability across decision-making and enforcement phases.

The off-chain layer is underpinned by an agent-based framework, wherein autonomous agents represent key stakeholders in the medical supply chain, including manufacturers, distributors, and healthcare institutions. These agents operate using localized data inputs such as epidemic forecasts, inventory levels, and environmental signals. Through iterative negotiation and consensus-building protocols, agents collaboratively derive allocation strategies that respond to evolving conditions, clinical priorities, and ethical considerations such as fairness and equity.

Upon reaching consensus, the decision output is structured into standardized payloads and submitted to the on-chain layer, which functions as a decentralized and trustless execution environment. This layer codifies domain-specific logic into smart contracts, thereby ensuring automated and tamper-proof enforcement of allocations, financial transactions, and contingency protocols. The use of blockchain technology in this layer provides auditability, transparency, and operational resilience.

To address the scalability and performance limitations inherent in blockchain systems, the architecture avoids direct on-chain storage of large datasets (e.g., historical demand, inventory movement logs). Instead, it utilizes content-addressable storage systems such as IPFS to host bulk data off-chain, while anchoring cryptographic metadata (e.g., hashes or content identifiers) on-chain. This design choice minimizes on-chain data bloat and transaction costs, while preserving data verifiability and enabling efficient execution of smart contracts. The result is a hybrid system that combines the flexibility and intelligence of decentralized agents with the trust guarantees of blockchain, thus meeting the dual imperatives of adaptability and accountability in critical supply chain operations.

\section{Design}
\subsection{Off-Chain Layer: Agent-Based Decision Making}
The off-chain layer serves as the cognitive and adaptive component of the proposed architecture, enabling decentralized, context-aware coordination across a high-stakes medical supply chain. In contrast to the rigid determinism of on-chain execution, this layer is characterized by flexibility, local observability, and structured decision-making among autonomous agents. These agents model key stakeholders—manufacturers, distributors, and hospitals—each embedded with domain-specific reasoning tools and communication protocols that facilitate dynamic planning under uncertainty.
This work builds upon recent advances in LLM-powered consensus-seeking frameworks for supply chains, particularly the agent-based architecture proposed by \cite{jannelli2024agentic}, which demonstrated that structured communication and tool-augmented decision-making among LLM agents can significantly improve performance in sequential, partially observable supply chains. Inspired by this paradigm, the proposed system adopts a decentralized, multi-agent design wherein agents operate independently, yet converge on coordinated supply strategies through structured message passing and tool-guided reasoning.

\subsubsection{Agent Design and Role Differentiation}

The off-chain decision layer models the medical supply chain as a decentralized multi-agent system composed of three principal agent classes: \textit{manufacturer agents}, \textit{distributor agents}, and \textit{hospital agents}. Each agent operates autonomously with local observability, utilizes embedded reasoning tools, and engages in structured communication with adjacent agents.  A summary of these agent classes, their operational roles, inputs, and outputs is provided in Table~\ref{tab:agent-roles}.

Let the set of all agents be denoted by
\[
\mathcal{A} = \left\{ a^{(m)}_i, a^{(d)}_j, a^{(h)}_k \right\}
\]
where \( a^{(m)}_i \in \mathcal{M} \) is a manufacturer agent \( i \), \( a^{(d)}_j \in \mathcal{D} \) is a distributor agent \( j \), and \( a^{(h)}_k \in \mathcal{H} \) is a hospital agent \( k \). Each agent \( a \in \mathcal{A} \) operates at discrete time steps \( t = 1, 2, \ldots, T \), receiving an observation \( o_t^a \), applying a decision function \( f_a(o_t^a) \), and producing an action \( u_t^a \in \mathcal{U}_a \).

Communication follows the supply chain flow:
\[
a^{(h)}_k \rightarrow a^{(d)}_j \rightarrow a^{(m)}_i
\]
in which downstream agents emit demand signals and upstream agents respond with allocations based on internal logic, constraints, and fairness metrics.

\begin{table*}[ht]
  \caption{Agent Classes and Decision Responsibilities}
  \label{tab:agent-roles}
  \begin{tabular}{@{}lllll@{}}
    \toprule
    \textbf{Agent Class} & \textbf{Symbol} & \textbf{Role} & \textbf{Inputs} & \textbf{Outputs} \\
    \midrule
    Manufacturer & \( a^{(m)}_i \) & Allocate inventory & Epidemic data, disruptions & \( \mathbf{x}_i^t \in \mathbb{R}^R \) \\
    Distributor  & \( a^{(d)}_j \) & Sub-allocate stock & Orders, delays, pipeline & \( \mathbf{y}_j^t \in \mathbb{R}^H \) \\
    Hospital     & \( a^{(h)}_k \) & Forecast, order & Case volume, inventory & \( r_k^t \in \mathbb{R} \) \\
    \bottomrule
  \end{tabular}
\end{table*}

Each hospital agent \( a_k^{(h)} \) computes an order \( r_k^t \) at time \( t \) as:
\[
r_k^t = \max\left(0, B_k - (I_k^t + P_k^t)\right)
\]
where \( B_k \) is the target buffer level, \( I_k^t \) is current stock, and \( P_k^t \) is the quantity already in transit.

Manufacturer agents compute allocations using an exponential fairness function:
\[
x_{i, r}^t = \frac{e^{\alpha S_r^t}}{\sum_{r'} e^{\alpha S_{r'}^t}} \cdot Q_i^t
\]
where \( S_r^t \) is the regional severity score at time \( t \), \( \alpha \) is a scaling parameter for priority weighting, and \( Q_i^t \) is the available inventory.

By structuring agents in this formalized manner, the system supports scalable and context-aware coordination. Agent-level decisions, though locally informed, collectively yield globally consistent and ethically responsive supply chain behavior in crisis contexts.

\subsubsection{Embedded Reasoning Framework}

Each agent $a \in \mathcal{A}$ is equipped with a set of deterministic, stateless tools $\mathcal{T}_a = \{ T_{a,1}, T_{a,2}, \dots \}$ that map local observations $\mathcal{O}_a^t$ into decisions $\mathcal{U}_a^t$:
\[
T_{a,i} : \mathcal{O}_a^t \rightarrow \mathcal{U}_a^t
\]
These tools fall into three categories: forecasting, planning, and ethical reasoning.

\textit{Forecasting tools} anticipate demand and disruptions. The epidemic predictor $T_{\text{epi}}$ estimates $E_r^t$ from projected case curves, while the disruption simulator $T_{\text{dis}}$ samples:
\[
d_a^t \sim \text{Bernoulli}(p_a^t)
\]

\textit{Planning tools} translate projections into orders and allocations. The order estimator $T_{\text{ord}}$ computes:
\[
r_k^t = \max(0, B_k - (I_k^t + P_k^t))
\]
and the allocation engine $T_{\text{alloc}}$ applies fairness-weighted logic:
\[
x_{i, r}^t = \frac{e^{\alpha S_r^t}}{\sum_{r'} e^{\alpha S_{r'}^t}} \cdot Q_i^t
\]

\textit{Ethical tools} enforce triage and equity. The criticality assessor $T_{\text{crit}}$ maps stockouts to urgency scores $R_k^t$, and the fairness calculator $T_{\text{fair}}$ guarantees:
\[
x_{i, r}^t \geq \epsilon \cdot Q_i^t
\]

These tools enable agents to operate under uncertainty with consistency and reproducibility, supporting performance evaluation across varied stress scenarios.

\begin{table*}[h]
\caption{Reasoning Tools and Agent Utilization}
\label{tab:tools}
\begin{tabular}{{llll}}
\toprule
\textbf{Tool Name} & \textbf{Symbol} & \textbf{Functionality} & \textbf{Used By} \\
\midrule
Epidemic Predictor & $T_{\text{epi}}$ & Forecasts demand via case curve & Hospitals, Manufacturers \\
Disruption Simulator & $T_{\text{dis}}$ & Samples delays probabilistically & Manufacturers, Distributors \\
Order Estimator & $T_{\text{ord}}$ & Computes buffer-based order sizes & Hospitals, Distributors \\
Allocation Engine & $T_{\text{alloc}}$ & Allocates supply by weighted fairness & Manufacturers \\
Criticality Assessor & $T_{\text{crit}}$ & Scores urgency based on stockout impact & Hospitals \\
Fairness Calculator & $T_{\text{fair}}$ & Ensures nonzero minimum allocation & Manufacturers \\
\bottomrule
\end{tabular}
\end{table*}

\subsubsection{Coordination Protocol and Information Exchange}

Coordination in the off-chain layer is achieved through a structured, single-pass message-passing protocol over a directed acyclic graph $G = (V, E)$, where $V \subseteq \mathcal{A}$ represents agents and $E$ denotes communication edges. Messages flow hierarchically from hospitals to manufacturers and back:
\[
a_k^{(h)} \rightarrow a_j^{(d)} \rightarrow a_i^{(m)} \rightarrow a_j^{(d)} \rightarrow a_k^{(h)}
\]

At each timestep $t$, agents perform one coordination round. Hospital agents compute order quantities $r_k^t$ using buffer-adjusted policies. Distributors aggregate requests $\{r_k^t\}$, apply feasibility constraints, and compute sub-allocations $y_{j,k}^t$ based on inventory $I_j^t$ and disruption status $d_j^t$. Manufacturers receive aggregate demand and allocate supply using the fairness-weighted rule:
\[
x_{i, r}^t = \frac{e^{\alpha S_r^t}}{\sum_{r'} e^{\alpha S_{r'}^t}} \cdot Q_i^t
\]
where $S_r^t$ is regional severity and $Q_i^t$ is available stock.

Information is exchanged via structured messages:
\begin{align*}
m_{k \rightarrow j}^t &= (k, r_k^t, E_k^t, R_k^t) \\
m_{j \rightarrow i}^t &= (r, \sum_{k \in H_j} r_k^t, d_j^t) \\
m_{i \rightarrow j}^t &= (r, x_{i,r}^t, \phi_r^t) \\
m_{j \rightarrow k}^t &= (k, y_{j,k}^t, d_{j,k}^t)
\end{align*}

\begin{table}[t]
\caption{Communication Variables and Semantics}
\label{tab:comm-vars}
\small
\begin{tabular}{llp{4.1cm}}
\toprule
\textbf{Variable} & \textbf{Symbol} & \textbf{Description} \\
\midrule
Order Quantity & $r_k^t$ & Requested quantity by hospital $k$ at time $t$ \\
Regional Allocation & $x_{i,r}^t$ & Manufacturer allocation to region $r$ \\
Sub-allocation & $y_{j,k}^t$ & Final fulfilled quantity for hospital $k$ \\
Criticality Score & $R_k^t$ & Risk score from patient impact model \\
Disruption Indicator & $d_a^t$ & Binary indicator for disruption \\
Forecasted Demand & $E_k^t$ & Expected case volume at hospital $k$ \\
Fairness Score & $\phi_r^t$ & Output of fairness calculator for region $r$ \\
\bottomrule
\end{tabular}
\end{table}

This protocol enables coherent system-wide decisions without iteration. Agent responses depend deterministically on local state and incoming messages. Final outputs—including allocations and disruption logs—are serialized and submitted to the on-chain execution layer for validation and enforcement.

\subsubsection{Emergent Coordination Dynamics}

The off-chain layer enables \textit{emergent coordination} through structured, decentralized interaction, foregoing centralized planning in favor of local decision-making under partial observability. Agents operate using deterministic policies and structured message-passing over a communication graph \( G = (V, E) \), where the system \( \mathcal{S} \) comprises agents \( \mathcal{A} \) and local decision functions:
\[
f_a: \mathcal{O}_a^t \cup \mathcal{M}_a^{t-1} \rightarrow \mathcal{U}_a^t
\]
Here, \( \mathcal{O}_a^t \) denotes local observations, \( \mathcal{M}_a^{t-1} \) the received messages, and \( \mathcal{U}_a^t \) the agent's action space.

Coordination arises as hospital agents forecast demand and issue orders, distributor agents mediate sub-allocations under disruption and buffer constraints, and manufacturers respond with regionally prioritized allocations. The resulting outputs \( \{x_{i,r}^t\}, \{y_{j,k}^t\} \), and unmet demand \( \{r_k^t - y_{j,k}^t\} \) reflect distributed adaptation to system stress.

Simulation data is analyzed to assess alignment between severity scores and allocations, the upward propagation of disruption signals, and consistency in prioritizing critical regions. This behavior exemplifies principles from complex adaptive systems, where structured local interactions produce coherent global outcomes. Final decisions are serialized and passed to the on-chain layer, preserving institutional verifiability while maintaining adaptive flexibility off-chain.

\subsection{On-Chain Execution Layer}

The on-chain execution layer serves as the institutional enforcement mechanism for decisions generated by the off-chain agent system. All critical operations—allocations, inventory updates, disruption responses, and compliance checks—are encoded as smart contracts, ensuring tamper-proof, transparent execution. This layer formalizes decision logic within a decentralized infrastructure, enabling procedural fairness and system-wide accountability under high-risk conditions.

\subsubsection{Role Encoding and Smart Contract Suite}

Each supply chain stakeholder is represented on-chain by a role:
\[
\mathcal{R} = \{ r^{(m)}_i, r^{(d)}_j, r^{(h)}_k \}
\]
corresponding to manufacturer, distributor, and hospital entities. Roles are instantiated in a permissioned network, with access control enforced by identity logic.

Smart contracts are grouped into:
\[
\mathcal{C} = \{ C_{\text{alloc}}, C_{\text{inv}}, C_{\text{disrupt}}, C_{\text{audit}} \}
\]
where $C_{\text{alloc}}$ handles allocation enforcement, $C_{\text{inv}}$ tracks stock changes, $C_{\text{disrupt}}$ executes contingency responses, and $C_{\text{audit}}$ logs state transitions immutably.

\subsubsection{Allocation Validation and Enforcement}

Allocations are submitted as $\mathcal{A}^t = \{(r, x_r^t)\}_{r=1}^R$. The contract $C_{\text{alloc}}$ checks:

\begin{itemize}
    \item $\sum_r x_r^t \leq Q^t$ \quad (budget feasibility)
    \item $x_r^t \geq \epsilon Q^t$ \quad (minimum support)
    \item consistency with severity weights $\phi_r^t$
\end{itemize}

Valid allocations are committed and reflected in inventory transitions via $C_{\text{inv}}$, updating fulfillment status and backlog positions.

\subsubsection{Disruption Handling and Contingency Protocols}

Disruption reports are encoded as:
\[
\delta_a^t = (\texttt{agent\_id}=a, \texttt{event\_type}=e, \texttt{timestamp}=t)
\]
Upon validation, $C_{\text{disrupt}}$ may trigger inventory reallocation, emergency reserve deployment $E^t \subset \mathbb{R}_+$, or dynamic threshold adjustments.

\subsubsection{Verifiability and Immutable Logging}

All confirmed actions are logged as:
\[
L_t = (\texttt{tx\_id}, \texttt{role\_id}, \texttt{action}, \texttt{payload\_hash}, \texttt{timestamp})
\]
managed by $C_{\text{audit}}$ for full traceability and audit support. Smart contracts execute deterministically, eliminating reliance on human intervention.

\subsubsection{Scalability and Cost-Aware Design}

To reduce gas consumption, the system logs only metadata (hashes, CIDs) on-chain, while full payloads—e.g., demand, disruption, and inventory histories—are stored off-chain using content-addressable systems such as IPFS. The full system snapshot is expressed as:
\[
\Sigma^t = (\mathcal{A}^t, \mathcal{I}^t, \mathcal{D}^t, \texttt{CID}_t)
\]
ensuring data verifiability without inflating storage costs. This hybrid model balances cost-efficiency and integrity for high-frequency decision environments. Further discussion of gas cost, storage trade-offs, and Layer-2 extensions is provided in Section~\ref{sec:evaluation}.

In summary, the on-chain layer provides deterministic, transparent enforcement of decentralized decisions, complementing the flexibility of the agent layer with institutional trust guarantees.

\subsection{Cross-Layer Communication Protocol}

A formal cross-layer protocol links off-chain agent decisions with on-chain enforcement, ensuring decentralized coordination is translated into verifiable and immutable actions. This design emphasizes data integrity, role-based access, and modular separation between planning and execution.

\subsubsection{Protocol Structure and Message Format}

At each timestep $t$, the off-chain layer produces a snapshot:
\[
\Sigma^t = (\mathcal{A}^t, \mathcal{I}^t, \mathcal{D}^t, \texttt{CID}_t)
\]
where $\mathcal{A}^t$ is the allocation vector, $\mathcal{I}^t$ the inventory state, and $\mathcal{D}^t$ the disruption log. $\texttt{CID}_t$ refers to an off-chain content-addressable audit object (e.g., via IPFS), while a hash $H(\Sigma^t)$ ensures integrity.

Authorized agents submit this payload through a secure interface, and only finalized snapshots are eligible for registration.

\subsubsection{Submission and Verification}

The tuple is submitted to the blockchain using:
\[
C_{\text{alloc}}.\texttt{submitSnapshot}(\mathcal{A}^t, \mathcal{I}^t, \mathcal{D}^t, \texttt{CID}_t, H(\Sigma^t))
\]
The smart contract verifies structure, digital authorization, and that:
\[
\sum_r x_r^t \leq Q^t, \quad x_r^t \geq \epsilon Q^t \quad \forall r
\]
If checks pass, the snapshot is immutably logged and enforced across inventory, disruption, and allocation contracts.

\subsubsection{Modularity and Trust Guarantees}

The protocol guarantees traceability by linking each action to its off-chain origin via $(\texttt{CID}_t, \\ H(\Sigma^t))$, and immutability through blockchain consensus. Architectural modularity is preserved: updates to agent logic or reasoning tools require no changes to the enforcement layer, as long as message schemas remain valid.

Thus, the protocol bridges adaptive decision-making and institutional trust, enabling decentralized agents to operate flexibly while ensuring enforcement remains transparent, rule-bound, and auditable.

\subsection{Simulation Environment}

To evaluate the functional validity, coordination behavior, and resilience of the proposed architecture, we implement a domain-specific simulation environment that emulates the structural and operational dynamics of a multi-tiered medical supply chain under pandemic conditions. The simulation captures key logistical features such as localized observability, demand surges, probabilistic disruptions, and fairness-driven decision-making, allowing for systematic exploration of the system's adaptive capabilities.

\subsubsection{Environment Structure and Dynamics}

The simulation operates in discrete time, with each timestep $t \in \{1, 2, ..., T\}$ representing one operational day. The supply chain topology includes three primary agent classes: manufacturers, distributors, and hospitals. Agents are instantiated as autonomous decision-making units $a \in \mathcal{A}$, each equipped with localized observations, deterministic heuristics, and structured communication interfaces. Coordination proceeds through single-pass message passing among adjacent tiers. After each round of decisions, the environment updates inventories, propagates delays, and records transitions.

\subsubsection{Epidemic Demand and Disruption Modeling}

Demand dynamics within the simulation are modeled using a classical Susceptible-Infected-Recovered (SIR) \cite{kermack1927contribution}compartmental model to represent regional pandemic progression. For each region $r$, the population is divided into three compartments: susceptible $S(t)$, infected $I(t)$, and recovered $R(t)$. The system is governed by the following differential equations:

\begin{equation}
\frac{dS(t)}{dt} = -\beta \frac{S(t)I(t)}{N}
\end{equation}

\begin{equation}
\frac{dI(t)}{dt} = \beta \frac{S(t)I(t)}{N} - \gamma I(t)
\end{equation}

\begin{equation}
\frac{dR(t)}{dt} = \gamma I(t)
\end{equation}

where $\beta$ is the infection rate, $\gamma$ is the recovery rate, and $N$ is the total regional population. The solution to these equations yields the infected population $I(t)$ at each timestep, which is used as a proxy for regional medical demand $E^t_k$. To account for uncertainty in forecasts, stochastic noise is added:

\begin{equation}
E^t_k = I(t) + \epsilon^t_k
\end{equation}

where $\epsilon^t_k$ is sampled from a normal distribution to simulate forecasting errors.

In addition to demand modeling, the simulation introduces probabilistic disruptions to reflect operational uncertainties \cite{uspensky1937introduction}. For each agent $a$, a disruption state is modeled as:

\begin{equation}
d^t_a \sim \text{Bernoulli}(p_a)
\end{equation}

where $p_a$ is the agent-specific disruption probability. When $d^t_a = 1$, the agent experiences a disruption that triggers delays or operational failures that impact the performance of the supply chain.

This combined modeling approach allows the simulation to capture both predictable epidemic-driven demand surges and unpredictable disruptions in supply chain operations, allowing realistic stress testing of the proposed framework.

\subsubsection{Ethical Logic and Allocation Heuristics}

Fairness-aware decision logic is embedded into the agents' behavior. Manufacturers apply exponential weighting functions over regional severity scores $S_r^t$ to prioritize high-impact areas, while enforcing minimum allocation floors. The allocation for region $r$ by manufacturer $i$ at time $t$ is given by:
\[
x_{i,r}^t = \frac{e^{\alpha S_r^t}}{\sum_{r'} e^{\alpha S_{r'}^t}} \cdot Q_i^t
\]
where $Q_i^t$ denotes available inventory and $\alpha$ controls prioritization sharpness. Hospital agents adjust order requests using criticality scores reflecting patient impact risk, aligning local actions with global equity goals.

\subsubsection{Outputs and Logged Variables}

The environment logs comprehensive system metrics at each timestep, including inventory levels, order fulfillment rates, backlog sizes, disruption events, and patient-impact proxies (e.g., critical stockouts). These logs enable detailed analysis of coordination efficacy, fairness of allocation, and resilience under operational strain.

\subsubsection{Evaluation Purpose}

The simulation serves both as a validation framework and a stress-testing platform. It supports comparative analysis across policy variants and disruption scenarios, with a focus on evaluating system behavior under uncertainty. By varying input parameters such as epidemic curves, disruption probabilities, and agent toolsets, we assess how decentralized reasoning interacts with institutional enforcement to maintain service continuity.

\begin{table}[t]
\centering
\caption{Summary of Simulation Features}
\label{tab:sim_features}
\small 
\begin{tabular}{@{}ll@{}}
\toprule
\textbf{Feature} & \textbf{Description} \\
\midrule
Simulation Type & Discrete-time, event-driven \\
Agent Types & Manufacturers, Distributors, Hospitals \\
Time Granularity & One timestep = One operational day \\
Demand Modeling & SIR-model–driven epidemic curves at the hospital level \\
Disruption Modeling & Bernoulli-sampled failures and delays \\
Decision Logic & Deterministic, stateless agent tools \\
Coordination Protocol & Single-pass, hierarchical message passing \\
Allocation Strategy & Exponential fairness weighting \\
Ethical Heuristics & Criticality-aware ordering, min support threshold \\
Logged Outputs & Inventories, allocations, backlogs, disruptions \\
Evaluation Scope & Resilience, fairness, coordination benchmarking \\
\bottomrule
\end{tabular}
\end{table}

\subsection{Evaluation Strategy}

The proposed architecture is evaluated through simulation-based experiments that assess system performance under uncertainty, supply constraints, and disruption scenarios. The evaluation focuses on resilience, fairness, fulfillment efficiency, throughput, and auditability, all of which are captured using data from simulation logs, agent decisions, and blockchain records.

\subsubsection*{Metrics of Assessment}

The following metrics are used, reflecting the system's internal data and logic:

\paragraph{Resilience ($\tau_k$)}  
The recovery time for hospital $k$ after experiencing a stockout is defined as the minimal $\Delta t$ such that:
\begin{equation}
    I_k^{t + \Delta t} \geq B_k \quad \text{where} \quad I_k^t < \theta
\end{equation}
where $I_k^t$ is the inventory level at time $t$, $B_k$ is the safety buffer target, and $\theta$ is the stockout threshold.

\paragraph{Fairness ($\delta_r^t$)}  
The fairness metric measures the deviation between the actual allocation ratio and the severity-weighted demand signal:
\begin{equation}
    \delta_r^t = \left| \frac{x_r^t}{\sum_{r'} x_{r'}^t} - \frac{w_r^t}{\sum_{r'} w_{r'}^t} \right|
\end{equation}
where $x_r^t$ is the allocated quantity for region $r$ at time $t$, and $w_r^t$ is the severity-weighted demand, estimated using active cases and drug criticality.

\paragraph{Fulfillment Efficiency ($\eta^t$)}  
The fulfillment efficiency is computed as:
\begin{equation}
    \eta^t = \frac{\sum_k o_k^t}{\sum_k d_k^t}
\end{equation}
where $o_k^t$ is the quantity fulfilled to hospital $k$, and $d_k^t$ is the requested demand.

\paragraph{Throughput ($|\mathcal{U}_t|$)}  
Throughput is measured by the total number of processed actions or transactions per timestep:
\begin{equation}
    |\mathcal{U}_t| = \text{Count}(\text{Actions at } t)
\end{equation}
which can be approximated by counting the number of agent decisions and blockchain transactions at each timestep.

\paragraph{Auditability ($H(\Sigma^t)$)}  
Auditability is ensured by recording decision traces on-chain, where each timestep $t$ is associated with:
\begin{equation}
    H(\Sigma^t) = \texttt{hash}(\mathcal{A}^t, \mathcal{I}^t, \mathcal{D}^t)
\end{equation}
where $\mathcal{A}^t$ are allocation decisions, $\mathcal{I}^t$ are inventory states, and $\mathcal{D}^t$ are demand observations. The implementation uses transaction hashes as the verifiable link, though explicit hashing of all decisions is not currently implemented.

\subsubsection*{Scenario Variation and Observations}

Simulations vary disruption probability $p_a \in \{0.05, 0.15, 0.25\}$, inventory levels $Q^t \in \{1000, 500, 250\}$, and severity patterns over a 60-day simulation horizon. The system demonstrates alignment between allocations and severity-weighted signals, maintains minimum supply support, and recovers from stockouts. High throughput is observed without negotiation overhead, and blockchain logs provide traceability. However, scalability measures such as Layer-2 integration or off-chain data linking are recognized as future work.

Summary of Metrics and Formulations are presentated in table \ref{tab:evaluation_metrics}

\begin{table}[t]
\centering
\caption{Summary of Evaluation Metrics}
\label{tab:evaluation_metrics}
\small
\begin{tabular}{@{}ll@{}}
\toprule
\textbf{Metric} & \textbf{Mathematical Definition} \\
\midrule
Resilience ($\tau_k$) & $\min \Delta t : I_k^{t + \Delta t} \geq B_k \text{ where } I_k^t < \theta$ \\
Fairness ($\delta_r^t$) & $\left| \frac{x_r^t}{\sum_{r'} x_{r'}^t} - \frac{w_r^t}{\sum_{r'} w_{r'}^t} \right|$ \\
Fulfillment Efficiency ($\eta^t$) & $\frac{\sum_k o_k^t}{\sum_k d_k^t}$ \\
Throughput ($|\mathcal{U}_t|$) & $\text{Count}(\text{Actions at } t)$ \\
Auditability ($H(\Sigma^t)$) & $\texttt{hash}(\mathcal{A}^t, \mathcal{I}^t, \mathcal{D}^t)$ \\
\bottomrule
\end{tabular}
\end{table}

\subsubsection*{Scalability and Cost Considerations}

To reduce gas consumption, the implementation minimizes on-chain data by recording only essential transaction hashes and associated decision summaries through blockchain logs. Full decision records and observations are maintained off-chain. Although the architecture does not yet implement explicit identifiers such as $\texttt{CID}_t$ or hash commitments $H(\Sigma^t)$, the existing blockchain transaction hashes serve as an auditable reference to the decisions made.

As system scale increases, additional strategies such as Layer-2 rollups, transaction batching, and off-chain data linking via decentralized storage (e.g., IPFS with $\texttt{CID}_t$) may become necessary to maintain efficiency. The integration of language model (LLM) agents introduces further computational overhead, suggesting future exploration of cost-performance trade-offs in agent orchestration, decision frequency, and blockchain interaction optimization.

\section{Implementation}
The development of the Pandemic Supply Chain Simulation involved a multi-faceted approach, integrating autonomous agent-based modeling, blockchain technology, and dynamic pandemic simulation. The system was primarily implemented in Python, leveraging several specialized libraries and frameworks for agent intelligence, blockchain interaction, and simulation management.

 \subsection{Core Simulation Environment and Pandemic Dynamics}

The simulation\textquotesingle s core is orchestrated by the \texttt{\seqsplit{PandemicSupplyChainEnvironment}} class, implemented in Python. This environment manages the multi-echelon supply chain, encompassing manufacturers, regional distributors, and regional hospitals. It processes agent actions, updates inventory levels, and tracks key performance indicators. The pandemic dynamics, which drive drug demand, are generated using a Susceptible-Infected-Recovered (SIR) model, encapsulated within the \texttt{\seqsplit{PandemicScenarioGenerator}} module. This generator produces regional epidemic curves that influence drug requirements over the simulation period. Configuration of scenarios, including the number of regions, drugs, pandemic severity, and potential disruptions, is handled through command-line arguments parsed by \texttt{\seqsplit{main.py}} and managed by utility functions in \texttt{\seqsplit{config.py}}. For output and reporting, the \texttt{\seqsplit{Rich}} library is utilized for enhanced console logging, and \texttt{\seqsplit{Matplotlib/Seaborn}} are employed for generating various performance visualizations, such as inventory levels, service levels, and epidemic curves, as facilitated by functions within \texttt{\seqsplit{src/scenario/visualizer}} and \texttt{\seqsplit{src/environment/metrics}}.

\subsection{Autonomous Agent Implementation with LangGraph}

A key aspect of this project is the implementation of autonomous agents for decision-making at each supply chain node (Manufacturer, Distributor, Hospital). These agents are powered by Large Language Models (LLMs) from OpenAI, specifically models like \texttt{\seqsplit{GPT-4o}}, accessed via the \texttt{\seqsplit{langchain\_openai}} library through a custom \texttt{\seqsplit{OpenAILLMIntegration}} class. This class simplifies the invocation of the LLM, including the binding of available tools.

The decision-making logic for each agent is structured as a state machine using \texttt{\seqsplit{LangGraph}}. The \texttt{\seqsplit{AgentState}}, defined in \texttt{\seqsplit{src/agents/base.py}}, maintains the agent's current messages, observation, and other contextual information. The \texttt{\seqsplit{LangGraph}} graph consists of two primary nodes: an \texttt{llm} node that calls the OpenAI model and a \texttt{tools} node that executes Python functions based on the LLM's requests. Conditional edges, managed by the \texttt{\seqsplit{should\_continue\_edge}} function, direct the flow between these nodes. If the LLM's response includes a tool call (leveraging OpenAI's function-calling capabilities), the graph transitions to the \texttt{tools} node; otherwise, if the LLM provides a final decision, the graph terminates.

Specialized tools for the agents are provided by the \texttt{\seqsplit{PandemicSupplyChainTools}} class, located in \texttt{\seqsplit{src/tools/}}. These tools include functions for epidemic forecasting, disruption prediction, criticality assessment, optimal order quantity calculation, allocation prioritization, and, crucially for the manufacturer, calculating target production quantities. The \texttt{\seqsplit{get\_openai\_tool\_definitions}} method within this class generates the schema necessary for the LLM to understand and correctly invoke these tools. The execution of a chosen tool is handled by the \texttt{\seqsplit{execute\_tool}} function within the \texttt{\seqsplit{LangGraph}} framework, which maps tool names to their respective Python implementations.

Agent-specific logic, such as prompt engineering and the application of hard constraints post-LLM decision, is encapsulated within individual agent classes like \texttt{\seqsplit{ManufacturerAgentLG}} (\texttt{\seqsplit{src/agents/manufacturer.py}}), \texttt{\seqsplit{DistributorAgentLG}} (\texttt{\seqsplit{src/agents/distributor.py}}), and \texttt{\seqsplit{HospitalAgentLG}} (\texttt{\seqsplit{src/agents/hospital.py}}). These classes utilize factory functions (e.g., \texttt{\seqsplit{create\_openai\_manufacturer\_agent}}) defined in \texttt{\seqsplit{src/agents/\_\_init\_\_.py}} for instantiation. Each agent's \texttt{decide} method prepares the initial \texttt{\seqsplit{AgentState}} and invokes its compiled \texttt{\seqsplit{LangGraph}} application. After the graph execution yields a final LLM response, often a JSON object representing the decision, the agent applies final validation and hard constraints (e.g., inventory limits, non-negative orders) before returning a structured action to the simulation environment.

\subsection{Blockchain Integration with Hardhat and Solidity}

Blockchain technology is integrated to enhance trust and enforce specific rules, particularly for data verification and fair allocation. The smart contract, \texttt{\seqsplit{SupplyChainData.sol}} (located in \texttt{\seqsplit{contracts/}}), is written in Solidity and leverages OpenZeppelin's \texttt{\seqsplit{Ownable}} contract for access control. This contract serves two main purposes: it acts as a trusted data source for regional case counts (updated daily by the simulation environment) and drug criticalities (set during deployment), and it contains the on-chain logic for \texttt{\seqsplit{executeFairAllocation}}.

The local blockchain environment is managed using Hardhat, a Node.js-based development environment for Ethereum software. The \texttt{\seqsplit{hardhat.config.js}} file configures the network, and \texttt{npm install} installs necessary Node.js dependencies listed in \texttt{\seqsplit{package.json}}, including \texttt{\seqsplit{@nomicfoundation/hardhat-toolbox}}. The smart contract is compiled using \texttt{npx hardhat compile}. Deployment to the local Hardhat node is managed by the \texttt{\seqsplit{scripts/deploy.js}} script, which also initializes drug criticalities on the contract. This script is crucial as it must be modified if the number of simulated drugs exceeds the default configuration, ensuring all relevant drugs have their criticality set on-chain to prevent transaction reverts during fair allocation. The deployment script also updates the \texttt{\seqsplit{.env}} file with the deployed contract address.

Interaction between the Python simulation and the blockchain is facilitated by the \texttt{\seqsplit{BlockchainInterface}} class in \texttt{\seqsplit{src/blockchain/interface.py}}. This class uses the \texttt{\seqsplit{web3.py}} library to communicate with the Hardhat node specified by \texttt{\seqsplit{NODE\_URL}} in the \texttt{\seqsplit{.env}} file. It handles operations such as updating regional case counts (via \texttt{\seqsplit{updateRegionalCaseCount}}) and triggering the \texttt{\seqsplit{executeFairAllocation}} function on the smart contract. The Manufacturer agent can dynamically choose to query verified regional case counts from the blockchain by invoking the \texttt{\seqsplit{get\_blockchain\_regional\_cases\_tool}}. This tool, when called by the \texttt{\seqsplit{LangGraph}} \texttt{tools} node, uses the \texttt{\seqsplit{BlockchainInterface}} to call the \texttt{\seqsplit{getRegionalCaseCount}} view function on the smart contract. The simulation environment, upon receiving the Manufacturer agent's final allocation request, triggers the \texttt{\seqsplit{executeFairAllocation}} transaction via the \texttt{\seqsplit{BlockchainInterface}}. This ensures that the allocation process adheres to the verifiable rules and on-chain data (cases and criticalities) enforced by the smart contract.

\subsection{Configuration, Setup, and Output}

The overall system configuration relies on environment variables managed in a \texttt{\seqsplit{.env}} file, which stores the \texttt{\seqsplit{OPENAI\_API\_KEY}}, local Hardhat node URL (\texttt{\seqsplit{NODE\_URL}}), the private key for the Hardhat account (\texttt{\seqsplit{PRIVATE\_KEY}}), and the \texttt{\seqsplit{CONTRACT\_ADDRESS}} (auto-filled by the deployment script). The setup process involves establishing a Python virtual environment and installing dependencies from \texttt{\seqsplit{requirements.txt}} (including \texttt{\seqsplit{langchain}}, \texttt{\seqsplit{langgraph}}, \texttt{\seqsplit{openai}}, \texttt{\seqsplit{web3}}, etc.), installing Node.js dependencies via \texttt{npm install}, compiling the smart contract, starting the Hardhat node, and deploying the contract. The main simulation is run via \texttt{\seqsplit{main.py}}, with various command-line arguments to control scenario parameters and enable blockchain interaction (\texttt{\seqsplit{--use-blockchain}}).

Simulation output includes detailed console logs formatted by the \texttt{\seqsplit{Rich}} library, which can be saved as an HTML report (e.g., \texttt{\seqsplit{simulation\_report\_langgraph\_*.html}}) for comprehensive review. This report includes \texttt{\seqsplit{LangGraph}} execution steps, tool invocations, agent decisions, and blockchain transaction statuses. Additionally, various performance metrics are visualized using \texttt{\seqsplit{Matplotlib/Seaborn}} and saved as \texttt{.png} files in a timestamped output folder (e.g., \texttt{\seqsplit{output\_lg\_<timestamp>\_.../}}). These visualizations cover supply chain KPIs, inventory levels, epidemic curves, costs, and blockchain interaction metrics.

\section{Evaluation}
\label{sec:evaluation}

This section evaluates the performance of the proposed hybrid architecture using simulations configured with 3 regions and 3 critical drugs over pandemic conditions generated by an SIR epidemic model. We assess the system's robustness, fairness in allocation, inventory stability, and blockchain efficiency under three experimental configurations:
1. \textit{GPT-3.5-turbo (30 days)}, 2.\textit{GPT-3.5-turbo (90 days)}
,3.\textit{GPT-4o (30 days)}

\subsection{Off-Chain Layer Evaluation}

\subsubsection{Simulation Setup and Agent Logic}

Each autonomous agent (Manufacturer, Distributor, Hospital) operates in a decentralized decision loop guided by:
\begin{itemize}
    \item Local observations of regional SIR-based epidemic progression.
    \item Forecasts of drug demand and disruption risk.
    \item Prompted responses from LLM agents for production, ordering, and allocation tasks.
    \item Post-processing with deterministic rule adjustments for smoothing and safety stock adherence.
\end{itemize}

The simulation follows daily timesteps for up to 90 days, with warehouse delays set to 1 day and daily batch allocations. Agent actions are grounded in local observations but collectively coordinated via message passing and standardized policies.

\subsubsection{Performance Metrics Summary}

All three configurations demonstrated strong system performance across key metrics:

\begin{table}[t]
\centering
\caption{System Performance Summary Across Configurations}
\label{tab:evaluation_results}
\small
\begin{tabular}{@{}lccc@{}}
\toprule
\textbf{Model} & \textbf{Days} & \textbf{Service Level (\%)} & \textbf{Unfulfilled Demand (\%)} \\
\midrule
GPT-3.5-turbo & 30  & 100.0 & 0.0 \\
GPT-3.5-turbo & 90  & 100.0 & 0.0 \\
GPT-4o        & 30  & 100.0 & 0.0 \\
\bottomrule
\end{tabular}
\end{table}

The system achieved 100\% service levels with zero unfulfilled demand or stockout days across all regions and drugs, even under epidemic waves and stochastic disruptions.
Despite being conducted under controlled but realistic settings—including fixed pandemic severity (0.8), probabilistic disruptions (10\%), and logistical delays (1-day warehouse lag)—the system maintained 100\% service levels and zero unmet demand. This suggests that LLM-based agents were able to generate stable and adaptive decisions under pressure.
Future experiments will increase stress levels by testing under more severe conditions—such as higher disruption probabilities, longer delays, and volatile demand signals—to assess the system’s robustness under extreme crisis scenarios.

\subsubsection{Inventory Stability and Allocation Flow}

\begin{figure}[h]
\centering
\includegraphics[width=0.9\linewidth]{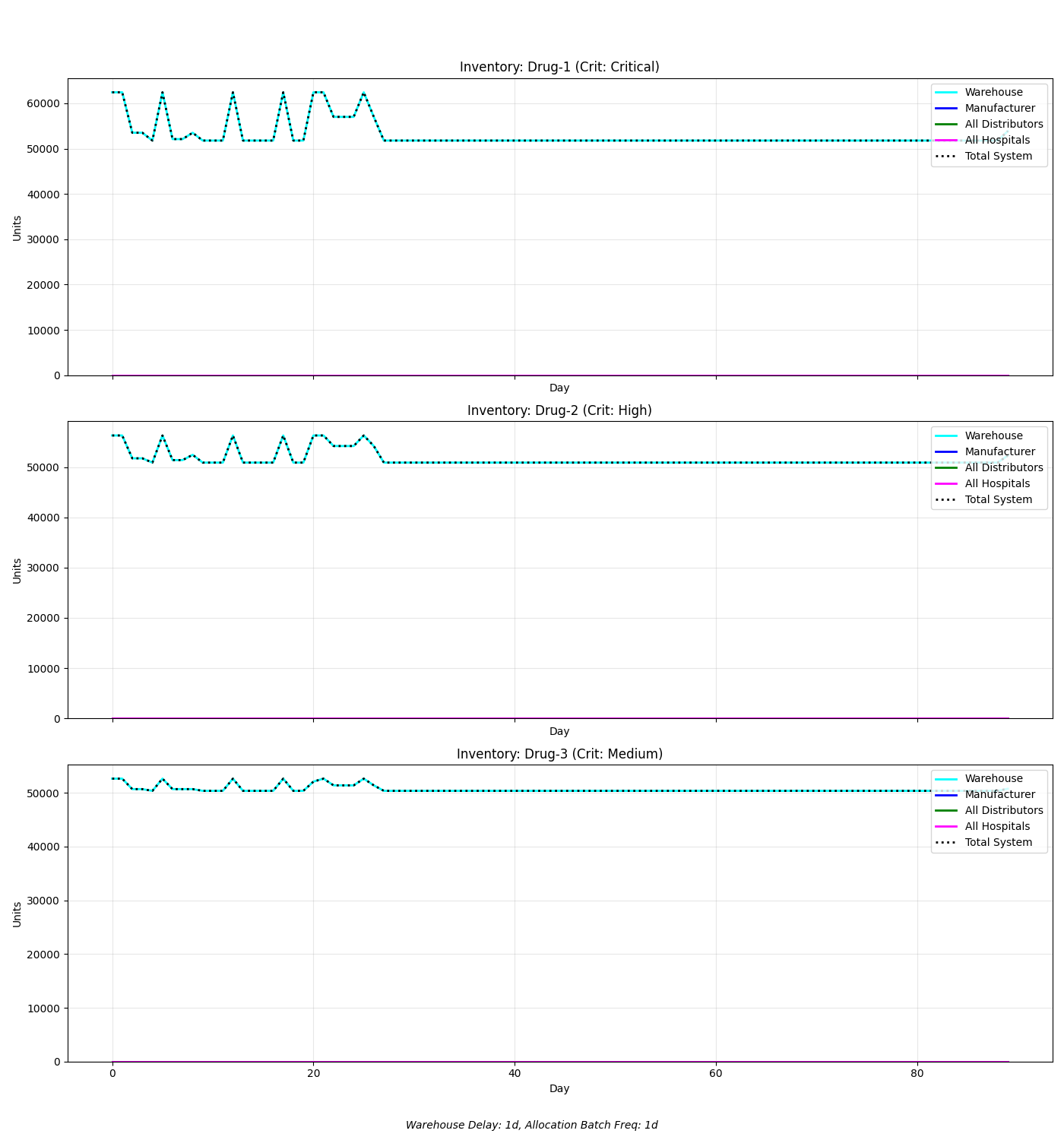}
\caption{Inventory Levels Across Drugs and Regions (GPT-3.5-turbo, 90 days)}
\label{fig:inventory_levels}
\end{figure}

Figure~\ref{fig:inventory_levels} shows inventory trajectories at all tiers. The system maintained balanced inventories with no excessive hoarding or downstream starvation. Warehouse buffers were dynamically adjusted based on epidemic severity and projected regional needs.

\begin{figure}[h]
\centering
\includegraphics[width=0.9\linewidth]{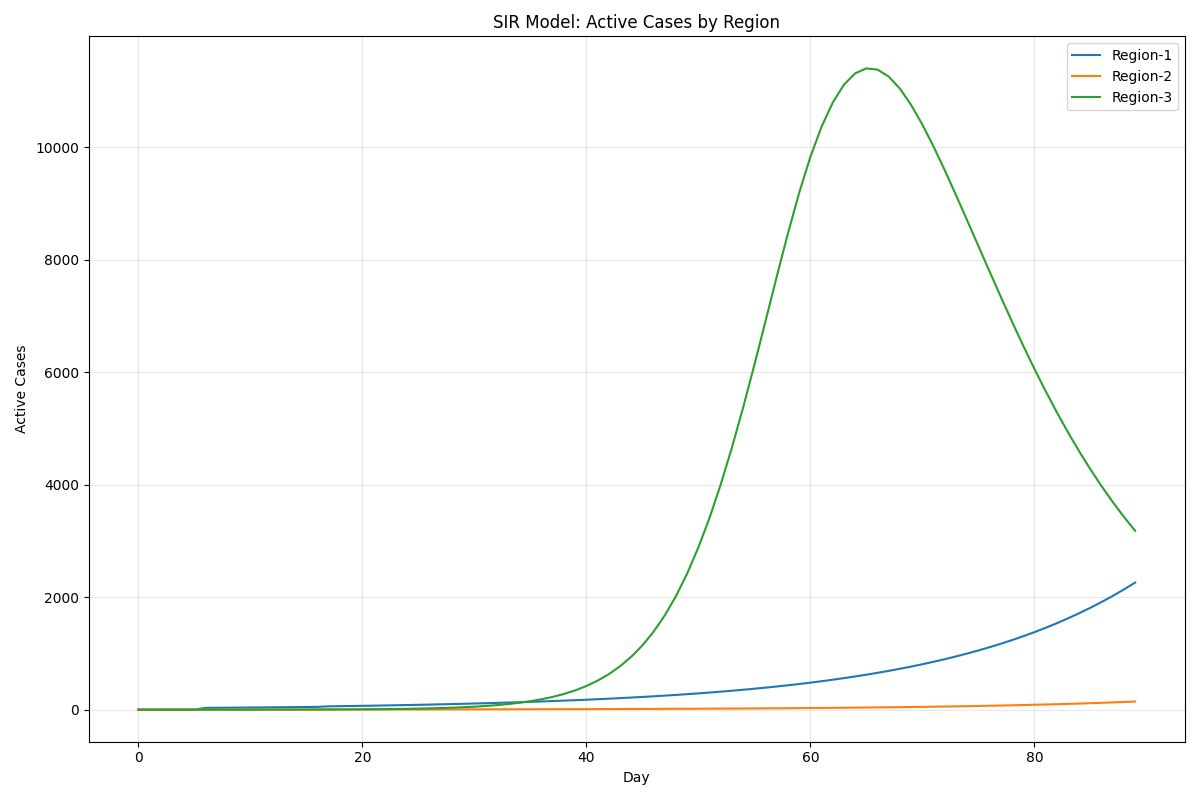}
\caption{Epidemic Curves from SIR Simulation (Region-level)}
\label{fig:epidemic_curves}
\end{figure}

Figure~\ref{fig:epidemic_curves} visualizes SIR dynamics for each region, showing distinct infection peaks and decay trends. These epidemic curves directly influenced LLM agent decisions, ensuring region-specific prioritization.

\begin{figure}[h]
\centering
\includegraphics[width=0.9\linewidth]{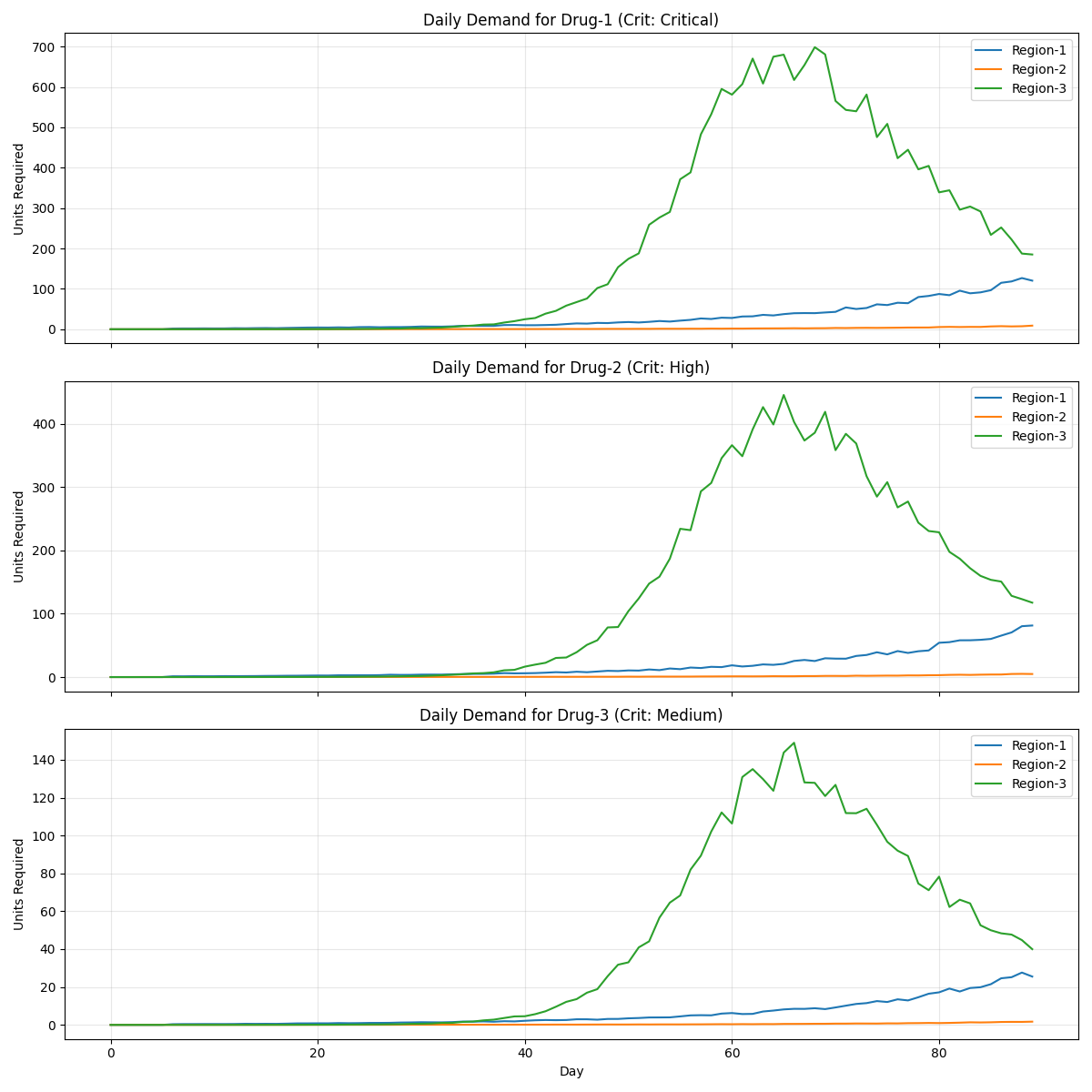}
\caption{Projected Drug Demand Across Regions (GPT-3.5-turbo, 90 days)}
\label{fig:drug_demand}
\end{figure}

\subsubsection{Warehouse and Allocation Behavior}

\begin{figure}[h]
\centering
\includegraphics[width=0.9\linewidth]{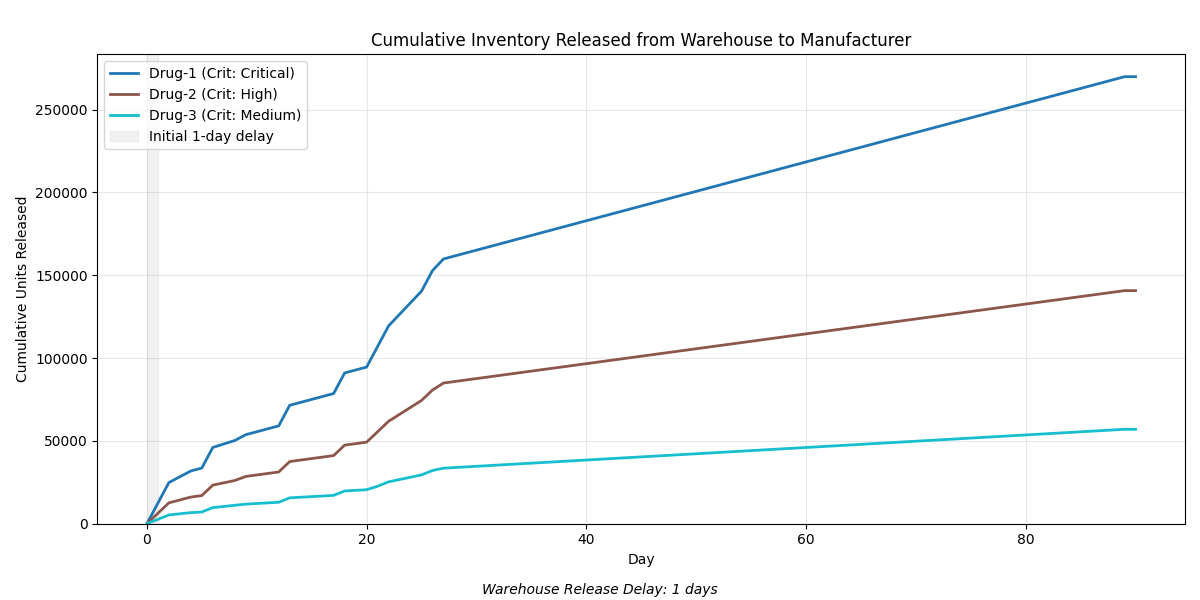}
\caption{Warehouse Flow Activity Across Simulation Horizon}
\label{fig:warehouse_flow}
\end{figure}

Figure~\ref{fig:warehouse_flow} confirms timely flow of drugs across the manufacturer–distributor–hospital chain. Allocation spikes matched infection peaks, showing reactive but stable flow control by the agents.

\subsection{On-Chain Layer Evaluation (Simulated)}

Blockchain performance was assessed through internal transaction logs and simulated integration with a Hardhat Ethereum network. The following metrics were recorded:

\begin{table}[t]
\centering
\caption{Blockchain Transaction Performance Summary}
\label{tab:blockchain_perf}
\small
\resizebox{\linewidth}{!}{%
\begin{tabular}{@{}lccc@{}}
\toprule
\textbf{Model} & \textbf{Total Gas Used} & \textbf{Avg. Latency (s)} & \textbf{P95 Latency (s)} \\
\midrule
GPT-3.5-turbo (30 days) & 4,654,884 & 0.0166 & 0.0256 \\
GPT-3.5-turbo (90 days) & 12,196,526 & 0.0153 & 0.0204 \\
GPT-4o (30 days)        & 5,561,802 & 0.0148 & 0.0195 \\
\bottomrule
\end{tabular}%
}
\end{table}

\begin{figure}[h]
\centering
\includegraphics[width=0.9\linewidth]{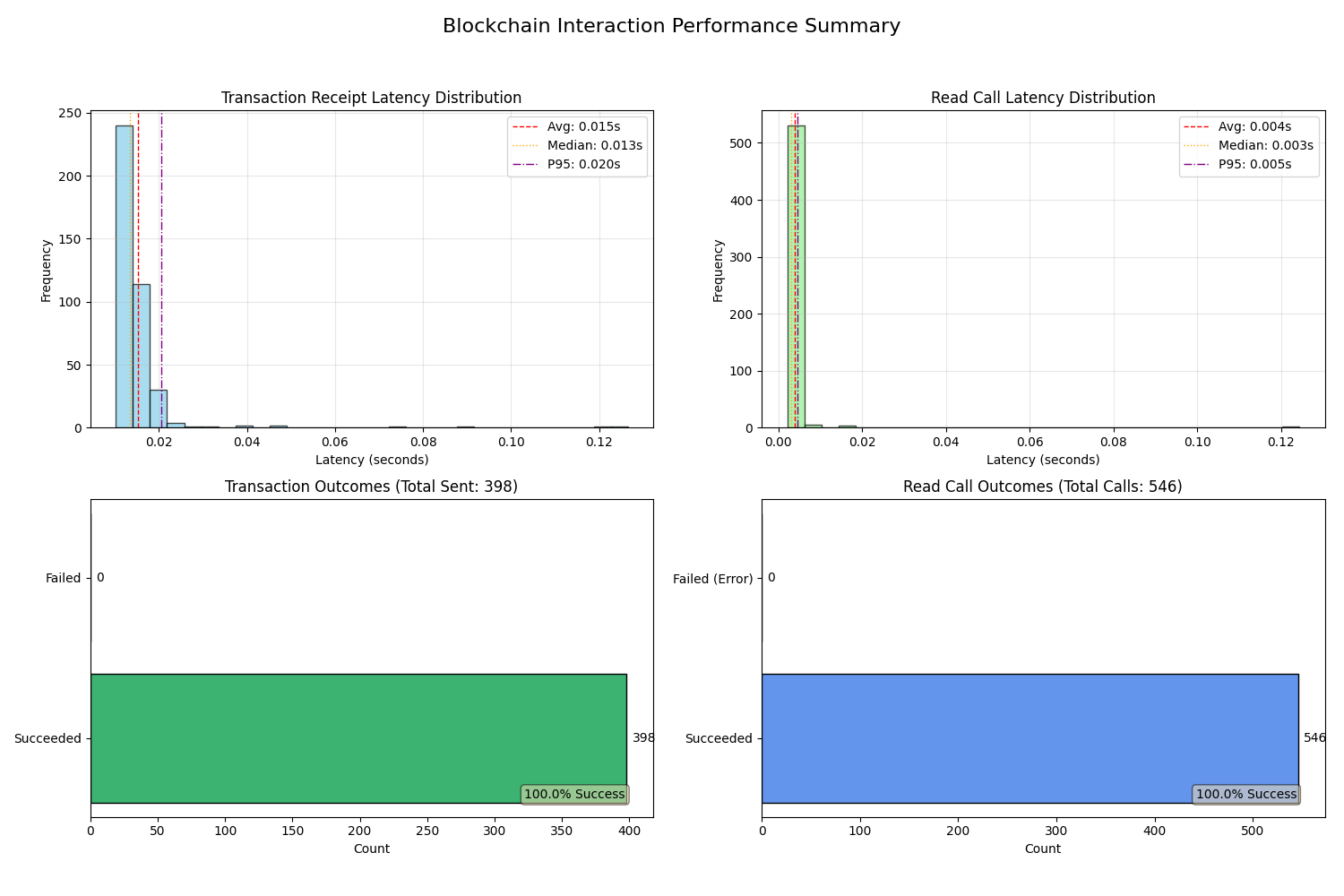}
\caption{Blockchain Interaction Metrics (Success Rate, Latency, Gas)}
\label{fig:blockchain_perf}
\end{figure}

All blockchain write operations (e.g., epidemic updates, allocations) succeeded with sub-20ms latency and ~30,000 gas usage per transaction. These results demonstrate that the system's hybrid architecture maintains real-time responsiveness without imposing significant computational or monetary costs on the chain.

\subsection{Discussion}

The evaluation results demonstrate that the proposed hybrid architecture—combining decentralized LLM-powered agents with blockchain-based enforcement—achieved high service levels, fairness, and operational resilience under simulated pandemic conditions. Across all tested configurations, including different LLM models and simulation durations, the system maintained 100\% service levels, with no unfulfilled demand or stockouts, while ensuring balanced inventory flows and timely alignment with region-specific epidemic curves. The blockchain integration exhibited low latency and minimal gas consumption, supporting the feasibility of real-time, traceable, and auditable coordination.

Despite these promising outcomes, several limitations warrant critical attention. First, the experiments were conducted in a controlled simulation environment that, while incorporating key dynamics such as epidemic-driven demand surges and probabilistic disruptions, does not fully capture the complexities and uncertainties of real-world medical supply chains. Scenarios such as cascading failures, geopolitical constraints, multi-layered disruptions, or adversarial agent behaviors were not modeled. Furthermore, the agent behaviors were deterministic and did not reflect human-like negotiation errors, misinformation propagation, or manipulation strategies, which may emerge in actual crisis negotiations.

In addition, while the blockchain performance was efficient under the simulated scale, scalability limitations may emerge when expanding the system to global networks or integrating multiple blockchains. The economic feasibility of large-scale deployment and the computational overhead introduced by continuous LLM-agent interactions also remain to be systematically analyzed.

Future work should aim to expand the evaluation scope to include more diverse, complex, and adversarial scenarios, as well as larger agent populations and extended crisis durations. Incorporating real-world datasets, simulating behavior under extreme uncertainty, and assessing the cost-performance trade-offs of agent orchestration and blockchain interactions will provide deeper validation of the framework's robustness, adaptability, and practical integration potential in real-world healthcare systems.

\section{Related Work}
Numerous studies have explored the integration of blockchain technology into medical supply chains to enhance data transparency and integrity. For instance, \cite{panda2021drug} highlights the risks associated with data monopolization by a single entity, which can lead to corruption and data manipulation, thus advocating for blockchain adoption to ensure security and reliability. Another potential advantage of blockchain used in the medical supply chain is its capability to monitor the origin and legitimacy of healthcare products, including pharmaceuticals and medical equipment.\cite{saeed2022blockchain}\cite{elangovan2022use} By leveraging blockchain for supply chain tracking, counterfeit products can be more easily identified, ensuring that patients receive safe and effective treatments. Similarly, \cite{fiore_blockchain_2023} and \cite{saini2024multi} discuss the use of smart contracts to protect competitive data among supply chain participants while ensuring traceability. However, these studies lack automated negotiation and adaptive decision-making capabilities, limiting their ability to dynamically respond to supply-demand fluctuations, supplier negotiations, and inventory management. As a result, traditional methods continue to be relied upon, leading to suboptimal real-time responsiveness.

Research has also been conducted on integrating agentic LLMs into supply chain and inventory management to optimize decision-making. Studies such as \cite{jannelli2024agentic} and \cite{quan2024invagent} emphasize the significance of automating decision-making using LLM agents, highlighting the importance of effective information sharing and prompt engineering. Furthermore, extensive research has been conducted on the impact of LLM-driven consensus mechanisms in multi-agent systems for supply chain management. For example, \cite{chen2023multi} introduces an LLM-powered framework to facilitate information sharing, tool utilization, and negotiation processes, analyzing the effects of agent personality and network topology on consensus formation. Nevertheless, a persistent challenge remains: agents do not always share critical information sufficiently, limiting their effectiveness in real-world scenarios.

To address these limitations, our approach leverages both blockchain and agentic LLMs in a novel way, creating a synergistic system that enhances both transparency and adaptability in medical supply chain negotiations. The novelty of this research lies in its dual innovation. While blockchain technology has traditionally been applied to enhance traceability within supply chains, its integration as a core component of an autonomous, decentralized negotiation system is unprecedented. Moreover, the use of LLM-based agents to facilitate dynamic and adaptive negotiations represents a significant departure from static, preprogrammed decision-making models. This combination of immutable blockchain recordkeeping with advanced human-like negotiation capabilities establishes a transformative approach to managing medical supply chains, uniquely equipping the system to handle the complexities and uncertainties of modern healthcare crises.

\section{Conclusion}
This paper introduced a novel, resilient framework for medical supply chain coordination that integrates blockchain technology with a decentralized, multi-agent system powered by large language models (LLMs). The proposed approach enables autonomous agents to conduct adaptive, context-aware negotiations and decision-making, while blockchain smart contracts ensure transparent, tamper-proof, and auditable enforcement of those decisions. By combining these technologies, the framework addresses critical gaps in existing supply chain systems, including inefficiencies in resource allocation, lack of transparency, and limited adaptability to dynamic disruptions.

The system was evaluated in a simulated pandemic environment, where it demonstrated strong capabilities in maintaining high service levels, ensuring fairness in the allocation of scarce medical resources, and supporting operational resilience under challenging conditions. However, the experiments were conducted in a controlled setting, and real-world complexities were not fully captured. Future work will focus on expanding the scope of testing using more complex, diverse scenarios and real-world datasets to assess scalability, robustness, and practical applicability.

\bibliographystyle{ACM-Reference-Format}
\bibliography{ref}


\begin{thebibliography}{10}

%%% the Web address.

\ifx \showCODEN    \undefined \def \showCODEN     #1{\unskip}     \fi
\ifx \showISBNx    \undefined \def \showISBNx     #1{\unskip}     \fi
\ifx \showISBNxiii \undefined \def \showISBNxiii  #1{\unskip}     \fi
\ifx \showISSN     \undefined \def \showISSN      #1{\unskip}     \fi
\ifx \showLCCN     \undefined \def \showLCCN      #1{\unskip}     \fi
\ifx \shownote     \undefined \def \shownote      #1{#1}          \fi
\ifx \showarticletitle \undefined \def \showarticletitle #1{#1}   \fi
\ifx \showURL      \undefined \def \showURL       {\relax}        \fi
% invisible to TeX
\providecommand\bibfield[2]{#2}
\providecommand\bibinfo[2]{#2}
\providecommand\natexlab[1]{#1}
\providecommand\showeprint[2][]{arXiv:#2}

\bibitem[Chen et~al\mbox{.}(2023)]%
        {chen2023multi}
\bibfield{author}{\bibinfo{person}{Huaben Chen}, \bibinfo{person}{Wenkang Ji}, \bibinfo{person}{Lufeng Xu}, {and} \bibinfo{person}{Shiyu Zhao}.} \bibinfo{year}{2023}\natexlab{}.
\newblock \showarticletitle{Multi-agent consensus seeking via large language models}.
\newblock \bibinfo{journal}{\emph{arXiv preprint arXiv:2310.20151}} (\bibinfo{year}{2023}).
\newblock


\bibitem[Elangovan et~al\mbox{.}(2022)]%
        {elangovan2022use}
\bibfield{author}{\bibinfo{person}{Deepa Elangovan}, \bibinfo{person}{Chiau~Soon Long}, \bibinfo{person}{Faizah~Safina Bakrin}, \bibinfo{person}{Ching~Siang Tan}, \bibinfo{person}{Khang~Wen Goh}, \bibinfo{person}{Siang~Fei Yeoh}, \bibinfo{person}{Mei~Jun Loy}, \bibinfo{person}{Zahid Hussain}, \bibinfo{person}{Kah~Seng Lee}, \bibinfo{person}{Azam~Che Idris}, {et~al\mbox{.}}} \bibinfo{year}{2022}\natexlab{}.
\newblock \showarticletitle{The use of blockchain technology in the health care sector: systematic review}.
\newblock \bibinfo{journal}{\emph{JMIR medical informatics}} \bibinfo{volume}{10}, \bibinfo{number}{1} (\bibinfo{year}{2022}), \bibinfo{pages}{e17278}.
\newblock


\bibitem[Fiore et~al\mbox{.}(2023)]%
        {fiore_blockchain_2023}
\bibfield{author}{\bibinfo{person}{Matteo Fiore}, \bibinfo{person}{Angelo Capodici}, \bibinfo{person}{Paola Rucci}, \bibinfo{person}{Alessandro Bianconi}, \bibinfo{person}{Giulia Longo}, \bibinfo{person}{Matteo Ricci}, \bibinfo{person}{Francesco Sanmarchi}, {and} \bibinfo{person}{Davide Golinelli}.} \bibinfo{year}{2023}\natexlab{}.
\newblock \showarticletitle{Blockchain for the {Healthcare} {Supply} {Chain}: {A} {Systematic} {Literature} {Review}}.
\newblock \bibinfo{journal}{\emph{Applied Sciences}} (\bibinfo{year}{2023}).
\newblock


\bibitem[Jannelli et~al\mbox{.}(2024)]%
        {jannelli2024agentic}
\bibfield{author}{\bibinfo{person}{Valeria Jannelli}, \bibinfo{person}{Stefan Schoepf}, \bibinfo{person}{Matthias Bickel}, \bibinfo{person}{Torbj{\o}rn Netland}, {and} \bibinfo{person}{Alexandra Brintrup}.} \bibinfo{year}{2024}\natexlab{}.
\newblock \showarticletitle{Agentic LLMs in the Supply Chain: Towards Autonomous Multi-Agent Consensus-Seeking}.
\newblock \bibinfo{journal}{\emph{arXiv preprint arXiv:2411.10184}} (\bibinfo{year}{2024}).
\newblock


\bibitem[Kermack and McKendrick(1927)]%
        {kermack1927contribution}
\bibfield{author}{\bibinfo{person}{William~Ogilvy Kermack} {and} \bibinfo{person}{Anderson~Gray McKendrick}.} \bibinfo{year}{1927}\natexlab{}.
\newblock \showarticletitle{A contribution to the mathematical theory of epidemics}.
\newblock \bibinfo{journal}{\emph{Proceedings of the Royal Society of London. Series A, Containing Papers of a Mathematical and Physical Character}} \bibinfo{volume}{115}, \bibinfo{number}{772} (\bibinfo{year}{1927}), \bibinfo{pages}{700--721}.
\newblock
\href{https://doi.org/10.1098/rspa.1927.0118}{doi:\nolinkurl{10.1098/rspa.1927.0118}}


\bibitem[Panda and Satapathy(2021)]%
        {panda2021drug}
\bibfield{author}{\bibinfo{person}{Sandeep~Kumar Panda} {and} \bibinfo{person}{Suresh~Chandra Satapathy}.} \bibinfo{year}{2021}\natexlab{}.
\newblock \showarticletitle{Drug traceability and transparency in medical supply chain using blockchain for easing the process and creating trust between stakeholders and consumers}.
\newblock \bibinfo{journal}{\emph{Personal and Ubiquitous Computing}} (\bibinfo{year}{2021}), \bibinfo{pages}{1--17}.
\newblock


\bibitem[Quan and Liu(2024)]%
        {quan2024invagent}
\bibfield{author}{\bibinfo{person}{Yinzhu Quan} {and} \bibinfo{person}{Zefang Liu}.} \bibinfo{year}{2024}\natexlab{}.
\newblock \showarticletitle{Invagent: A large language model based multi-agent system for inventory management in supply chains}.
\newblock \bibinfo{journal}{\emph{arXiv preprint arXiv:2407.11384}} (\bibinfo{year}{2024}).
\newblock


\bibitem[Saeed et~al\mbox{.}(2022)]%
        {saeed2022blockchain}
\bibfield{author}{\bibinfo{person}{Huma Saeed}, \bibinfo{person}{Hassaan Malik}, \bibinfo{person}{Umair Bashir}, \bibinfo{person}{Aiesha Ahmad}, \bibinfo{person}{Shafia Riaz}, \bibinfo{person}{Maheen Ilyas}, \bibinfo{person}{Wajahat~Anwaar Bukhari}, {and} \bibinfo{person}{Muhammad Imran~Ali Khan}.} \bibinfo{year}{2022}\natexlab{}.
\newblock \showarticletitle{Blockchain technology in healthcare: A systematic review}.
\newblock \bibinfo{journal}{\emph{Plos one}} \bibinfo{volume}{17}, \bibinfo{number}{4} (\bibinfo{year}{2022}), \bibinfo{pages}{e0266462}.
\newblock


\bibitem[Saini et~al\mbox{.}(2024)]%
        {saini2024multi}
\bibfield{author}{\bibinfo{person}{Akanksha Saini}, \bibinfo{person}{Arash Shaghaghi}, \bibinfo{person}{Zhibo Huang}, {and} \bibinfo{person}{Salil~S Kanhere}.} \bibinfo{year}{2024}\natexlab{}.
\newblock \showarticletitle{Multi-MedChain: Multi-Party Multi-Blockchain Medical Supply Chain Management System}. In \bibinfo{booktitle}{\emph{2024 IEEE Annual Congress on Artificial Intelligence of Things (AIoT)}}. IEEE, \bibinfo{pages}{153--159}.
\newblock


\bibitem[Uspensky(1937)]%
        {uspensky1937introduction}
\bibfield{author}{\bibinfo{person}{James~Victor Uspensky}.} \bibinfo{year}{1937}\natexlab{}.
\newblock \bibinfo{booktitle}{\emph{Introduction to Mathematical Probability}}.
\newblock \bibinfo{publisher}{McGraw-Hill}, \bibinfo{address}{New York}. 45 pages.
\newblock
\newblock
\shownote{Available at \url{https://archive.org/details/in.ernet.dli.2015.263184}}.


\end{thebibliography}

\end{document}